\documentclass[12pt]{article}
\pdfoutput=1 
\usepackage[dvipdfmx]{graphicx}
\usepackage{amsfonts}
\usepackage[mathscr]{eucal}
\usepackage{ascmac}
\usepackage{dcolumn}
\usepackage{bm}
\usepackage[colorlinks=true,linkcolor=blue,citecolor=blue]{hyperref}
\usepackage{color}
\usepackage{amsmath}

\begin{document}


\newcommand{\gtrsim}{ \mathop{}_{\textstyle \sim}^{\textstyle >} }
\newcommand{\lesssim}{ \mathop{}_{\textstyle \sim}^{\textstyle <} }
\newcommand{\vev}[1]{ \left\langle {#1} \right\rangle }
\newcommand{\bra}[1]{ \langle {#1} | }
\newcommand{\ket}[1]{ | {#1} \rangle }
\newcommand{\EV}{ \ {\rm eV} }
\newcommand{\KEV}{ \ {\rm keV} }
\newcommand{\MEV}{\  {\rm MeV} }
\newcommand{\GEV}{\  {\rm GeV} }
\newcommand{\TEV}{\  {\rm TeV} }
\newcommand{\1}{\mbox{1}\hspace{-0.25em}\mbox{l}}
\newcommand{\Red}[1]{{\color{red} {#1}}}

\newcommand{\lmk}{\left(}  
\newcommand{\rmk}{\right)}
\newcommand{\lkk}{\left[}  
\newcommand{\rkk}{\right]}
\newcommand{\lhk}{\left \{ }  
\newcommand{\rhk}{\right \} }
\newcommand{\del}{\partial}  
\newcommand{\la}{\left\langle} 
\newcommand{\ra}{\right\rangle}
\newcommand{\half}{\frac{1}{2}}

\newcommand{\bea}{\begin{array}}
\newcommand{\eea}{\end{array}}
\newcommand{\beq}{\begin{eqnarray}}
\newcommand{\eeq}{\end{eqnarray}}

\newcommand{\dd}{\mathrm{d}}
\newcommand{\Mpl}{M_{\rm Pl}}
\newcommand{\mg}{m_{3/2}}
\newcommand{\abs}[1]{\left\vert {#1} \right\vert}
\newcommand{\mphi}{m_{\phi}}
\newcommand{\Hz}{\ {\rm Hz}}
\newcommand{\for}{\quad \text{for }}
\newcommand{\Min}{\text{Min}}
\newcommand{\Max}{\text{Max}}
\newcommand{\Kahler}{K\"{a}hler }
\newcommand{\cphi}{\varphi}

\begin{titlepage}

\baselineskip 8mm

\begin{flushright}
IPMU 17-0026
\end{flushright}

\begin{center}

\vskip 1.2cm

{\Large\bf New-Type Charged Q-ball Dark Matter in Gauge Mediated SUSY Breaking Models
}

\vskip 1.8cm

{\large 
Jeong-Pyong Hong$^{a,b}$ and Masahiro Kawasaki$^{a,b}$
}

\vskip 0.4cm

{\it$^a$Institute for Cosmic Ray Research, The University of Tokyo,
5-1-5 Kashiwanoha, Kashiwa, Chiba 277-8582, Japan}\\
{\it$^b$Kavli IPMU (WPI), UTIAS, The University of Tokyo, 5-1-5 Kashiwanoha, 
Kashiwa, 277-8583, Japan}

\date{\today}
\vspace{2cm}

\begin{abstract}
We examined the viability of new-type charged Q-ball dark matter scenario. We considered the case where the relics can be treated as ordinary 
ions, to which we applied the observational constraint from MICA experiment, where no trail of heavy 
ion-like object is observed in $10^9$ years old ancient muscovite mica. We have found that the allowed parameter region exists but is smaller than the neutral Q-ball dark matter case. We have also discussed whether our scenario is actually consistent with Affleck-Dine mechanism.
\end{abstract}


\end{center}
\end{titlepage}

\baselineskip 6mm


\section{Introduction
\label{sec:introduction}}
Affleck-Dine mechanism~\cite{ad} is one of promising candidates for baryogenesis based on supersymmetric (SUSY) theories, where baryon asymmetry is generated by the dynamics in the phase direction of a baryonic scalar field~(called Affleck-Dine field) such as squark. After the baryon number generation, the spatial inhomogeneities of the Affleck-Dine field due to quantum fluctuations grows exponentially into non-topological solitons, which are called Q-balls~\cite{ks,kkins,kkins2}. Q-ball is defined as a spherical solution in a global $U(1)$ theory which minimizes the energy of the system with fixed $U(1)$ charge~\cite{c}. The baryon number generated in Affleck-Dine mechanism is confined inside Q-balls. Although Q-balls are stable against decay into squarks, they may gradually decay into quarks and/or leptons, so that the baryon asymmetry in the universe is generated by quarks emitted from the Q-balls. However, in gauge mediated SUSY breaking models, a baryonic Q-ball can be stable against decay into nucleons~\cite{dk}, while a leptonic Q-ball can decay into leptons.

In the previous work~\cite{hyk}, we focused on gauge mediation type Q-balls that carry both baryon and lepton charges which can be formed after the Affleck-Dine baryogenesis with $u^cu^cd^ce^c$ flat direction, for instance
. We found that the Q-balls can be electrically charged, which are called charged or gauged Q-balls~\cite{gaugedqball}, and the charged Q-balls are stable by virtue of the stability of the baryonic component. Thus, the charged Q-ball can be the dark matter in the present universe. Since the charged Q-balls capture the charged particles, they form 
ion-like objects which make different experimental signatures compared to the ordinary neutral Q-balls. For instance, Q-balls can be detected by Super-Kamiokande~\cite{kensyutu, kttn} or IceCube~\cite{icecube}, which probe KKST process~\cite{sipal} where the Q-ball absorbs quarks and emits pions of energy~$\sim1\mathrm{GeV}$. However, the charged Q-ball relics experience the ordinary electromagnetic processes as well, so that the detection of those processes is applied. Numerous experiments and upper bounds on the flux of the relics from the charged Q-balls are given in Ref.~\cite{kensyutu}, and the most stringent constraint comes from MICA experiment~\cite{mica}, where the trail of heavy 
ion-like object is not observed in $10^9$ years old ancient mica crystals. Since the observation time is equal to the age of the mica, the constraint on Q-ball flux is much severer than those coming from other experiments. In the previous work~\cite{hyk2}, we found that the MICA constraint is more stringent than that from IceCube, etc. and allowed region of the model parameters for gauge mediation type charged Q-ball dark matter becomes smaller than the neutral Q-ball case. 

While the gauge mediation type Q-ball is realized when the gauge mediation effect dominates the potential of the Affleck-Dine field, gravity mediation effect may still dominate over the gauge mediation effect if the field value is sufficiently large. In that case, Q-balls with different properties are formed, which we call new-type~(or hybrid) Q-balls. In this paper, we assume that the new-type Q-balls are formed, instead of the gauge mediation type Q-balls, after the Affleck-Dine mechanism, and investigate whether there exists the allowed parameter region for the new-type charged Q-ball dark matter. In specific, we apply the MICA constraint on the relics from new-type charged Q-balls and find the constraint on gravitino mass~(SUSY breaking scale) and reheating temperature.
\section{New-type Q-ball}
In MSSM, there are numerous flat directions such as $u^cd^cd^c$, and $u^cu^cd^ce^c$, for instance. We consider dynamics of one of such flat directions in the MSSM. In particular, we consider a flat direction with nonzero baryon and lepton charges, which we call the Affleck-Dine field.

In our universe, the supersymmetry is broken, so the flat directions obtain soft SUSY breaking terms. We consider the minimal gauge mediation model~\cite{mgm,mg2,mg3}, where SUSY is spontaneously broken by F-term of a singlet field Z:
\begin{align}
\langle F_Z\rangle=F\neq0.
\end{align}
The soft breaking effect is mediated to the visible sector by messenger fields $\Psi$ and $\bar{\Psi}$, which is a pair of some representations and anti-representations of the minimal GUT group $SU(5)$, via the following interactions:
\begin{align}
W=kZ\bar{\Psi}\Psi+M_{\text{mess}}\bar{\Psi}\Psi,
\end{align}
where $k$ and $M_{\mathrm{mess}}$ denote a yukawa constant and messenger mass, respectively.

Then, the Affleck-Dine field obtains the following potential~\cite{ks,mmgc,17} by the breaking effect above:
\begin{align}
V=V_{\text{gauge}}+V_{\text{grav}}=M_F^4\left(\log\left(\frac{\left|\Phi\right|^2}{M_{\text{mess}}^2}\right)\right)^2+m_{3/2}^2\left(1+K\log\left(\frac{\left|\Phi\right|^2}{M_{\ast}^2}\right)\right)\left|\Phi\right|^2+(\text{A-term}).
\end{align}
The first term comes from gauge mediation effect and gauge mediation parameter $M_F$ is defined by
\begin{align}
M_F\simeq\frac{\sqrt{gkF}}{4\pi}
\label{eq:mfdef}
\end{align}
where $g$ generically denotes the gauge coupling of the standard model~\cite{mmgc}. 
The second term originates from gravity mediation effect and the gravitino mass $m_{3/2}$ is given as 
\begin{align}
m_{3/2}\equiv\frac{F}{\sqrt{3}M_{\text{P}}},
\label{eq:grma}
\end{align}
where $M_P=2.4\times10^{18}~\mathrm{GeV}$ is the reduced Planck mass. Here we assume $m_{3/2}<1\text{GeV}$. The parameter $K$ is a constant which comes from the beta function of mass of the AD field and is typically negative, satisfying $0.01\lesssim |K|\lesssim0.1$, and $M_*$ means the renormalization scale. The third term~(A-term)~is a $B$ and $CP$ violating term which induces the Affleck-Dine rotation. 

During inflation, the AD field additionally obtain a negative Hubble induced mass term $-c_HH^2|\Phi|^2$ due to the coupling to the inflaton, which gives AD field a large vacuum expectation value~(VEV). However, after inflation, the Hubble rate becomes smaller than the soft mass and the AD field starts the oscillation around the origin. Then, the spatial inhomogeneities of the AD field, which originate from quantum fluctuations, grow exponentially and form Q-balls. As mentioned in the previous section, new-type Q-ball~\cite{21} is formed if $V_{\text{grav}}$ dominates the potential when the AD field starts the oscillation. The condition for the $V_{\text{grav}}$ domination is given by
\begin{align}
\phi_{\text{osc}}>\phi_{\text{eq}}\simeq\sqrt{2}M_F^2/m_{3/2}
\label{eq:gad}
\end{align}
where $\phi\equiv\sqrt{2}|\Phi|$, and $\phi_{\text{osc}}$ is the field value at the beginning of the oscillation. A typical charge of new-type Q-ball can be estimated by a linear approximation or numerical simulations, which is given by~\cite{21}
\begin{align}
Q=\beta\left(\frac{\phi_{\text{osc}}}{m_{3/2}}\right)^2.
\label{eq:typicalcharge}
\end{align}
Here $\beta=0.02$, which is a numerical constant. 

The profile of the new-type Q-ball solution is known to be gaussian as
\begin{align}
\phi\propto \exp(-r^2/R_Q^2)
\label{eq:gau}
\end{align}
and mass, size and mass per unit charge of new-type Q-ball are given by
\begin{align}
M_Q&\simeq m_{3/2}Q\label{eq:mass}\\
R_Q&\simeq |K|^{-1/2}m_{3/2}^{-1}\label{eq:sizq}\\
\frac{dM_Q}{dQ}&\simeq m_{3/2},
\label{eq:omq}
\end{align}
respectively~\cite{23,24}.

In the usual scenario, the Q-ball is assumed to consist only baryonic components such as $u^cd^cd^c$, but here if we consider the simple two scalar model which consists of baryonic and leptonic components, whose more realistic case is $u^cu^cd^ce^c$, for example. Then, we can see that the baryonic component is stable against decay since $m_{3/2}<m_p$, and only the leptonic component can annihilate into $e^+$ inside Q-balls via gaugino and/or higgsino exchange interactions.
From this process, electric charge is induced in the Q-ball, and from the Gauss' law, we see that $U(1)_{\text{EM}}$ gauge field must have a classical configuration. In the next section, we consider the charged, or gauged Q-ball~\cite{gaugedqball}, which consists of not only the two scalars  but also the $U(1)$ gauge field. 
\section{New-type charged Q-ball}
\subsection{Setup and basic properties}
\label{sec:setp}
The Lagrangian of our two scalar model is written as
\begin{align}
&\mathcal{L}=(D_\mu\Phi_B)^\ast D^\mu\Phi_B+(D_\mu\Phi_L)^\ast D^\mu\Phi_L-V(\Phi_B,\Phi_L)-\frac14F_{\mu\nu}F^{\mu\nu},\\
&D_\mu\Phi_B=(\partial_\mu-ieA_\mu)\Phi_B,\\
&D_\mu\Phi_L=(\partial_\mu+ieA_\mu)\Phi_L,
\end{align}
and baryon and lepton charges are
\begin{align}
&B=\frac1{i}\int d^3x(\Phi_B^\ast D_0\Phi_B-\Phi_B(D_0\Phi_B)^\ast)\equiv\int d^3xb,\\
&L=\frac1{i}\int d^3x(\Phi_L^\ast D_0\Phi_L-\Phi_L(D_0\Phi_L)^\ast)\equiv\int d^3xl
\end{align}
where $b$ and $l$ are baryon and lepton number densities.
We assign the positive and negative electric charges 
for $B$ and $L$ components, respectively, and thus the total electric charge is given by 
\begin{align}
Q_e=B-L.
\end{align}
First, we set an ansatz on the configurations as follows\footnote{This parametrization is analogous to that of ordinary Q-ball, which is determined by the minimization of the energy for a fixed $U(1)$ charge, the definition of the Q-ball solution.}:
\begin{align}
\Phi_i&\equiv\frac1{\sqrt2}\phi_i,\ \ \   i=B, L\\
\phi_B(x,t)&=e^{i\omega_Bt}\phi_B(r),\\
\phi_L(x,t)&=e^{i\omega_Lt}\phi_L(r),\\
A_i&=0,\\
A_0&=A_0(r).
\end{align}

\begin{figure}[t]
\begin{minipage}{.45\linewidth}
  \includegraphics[width=\linewidth]{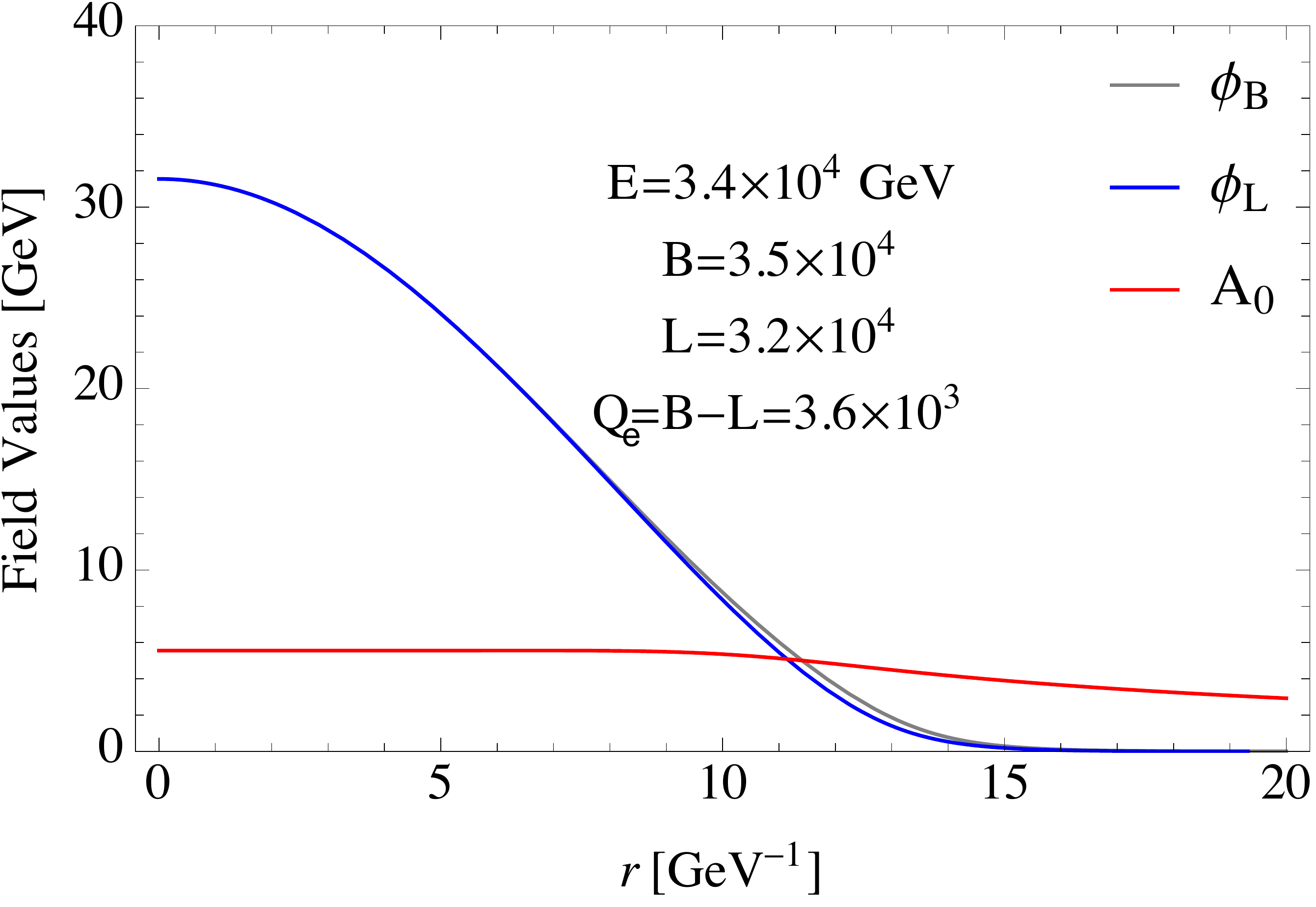}
 \end{minipage}
 \hspace{1cm}
 \begin{minipage}{.45\linewidth}
  \includegraphics[width=\linewidth]{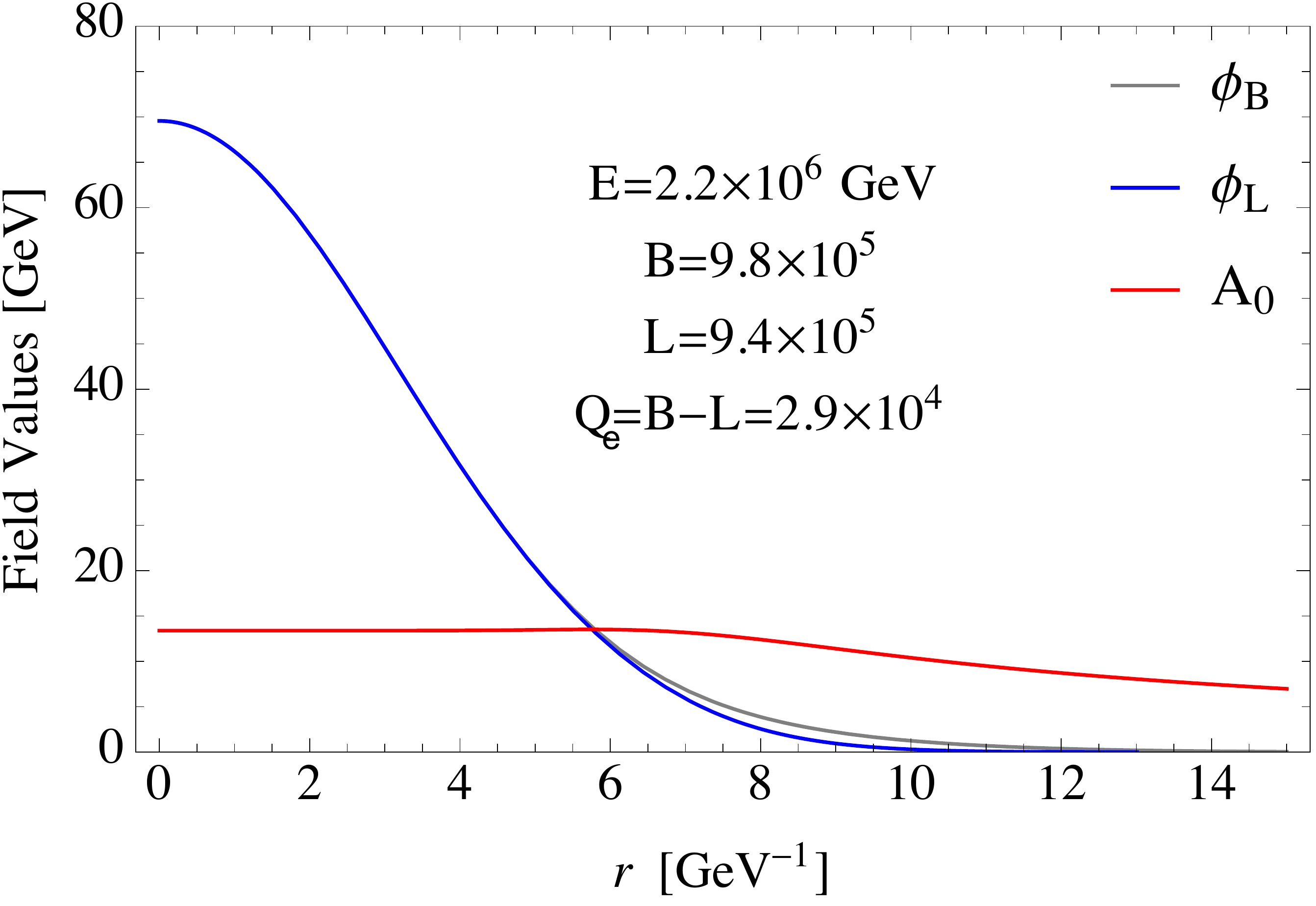}
 \end{minipage}
 \caption{Examples of profiles of charged Q-ball which has baryonic and leptonic components for gauge mediation type charged Q-ball, with $m_\phi=1\text{GeV}$, and $e^2=0.002$~(Left), and new-type charged Q-ball, with $m_{3/2}=1\text{GeV}$, $|K|=0.1$, and $e^2=0.002$~(Right). They are slightly deformed from the neutral profiles, due to small nonzero electric charge. }
\label{fig:prof}
\end{figure}

\begin{figure}[t]
\begin{minipage}{.45\linewidth}
  \includegraphics[width=\linewidth]{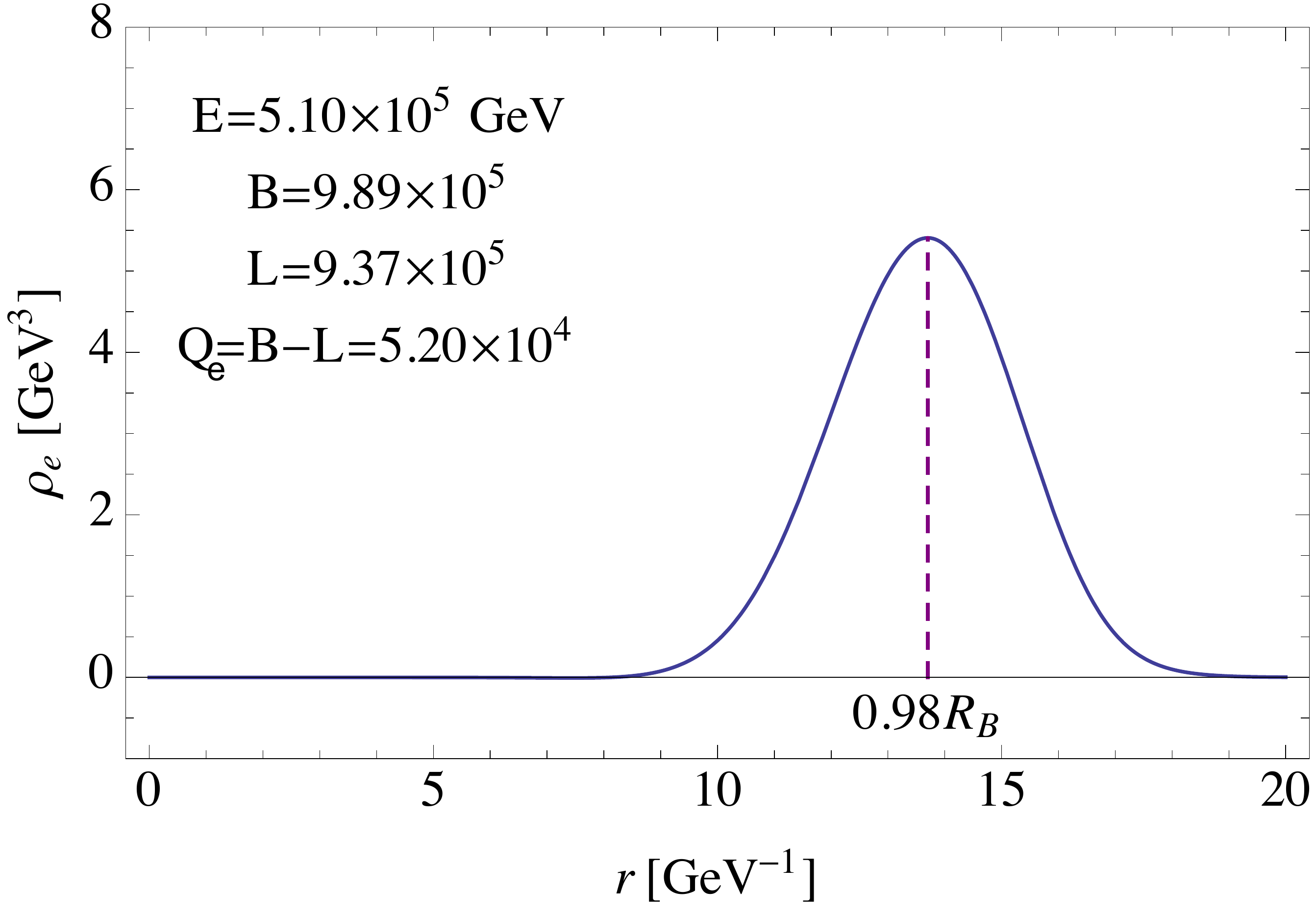}
 \end{minipage}
 \hspace{1cm}
 \begin{minipage}{.45\linewidth}
  \includegraphics[width=\linewidth]{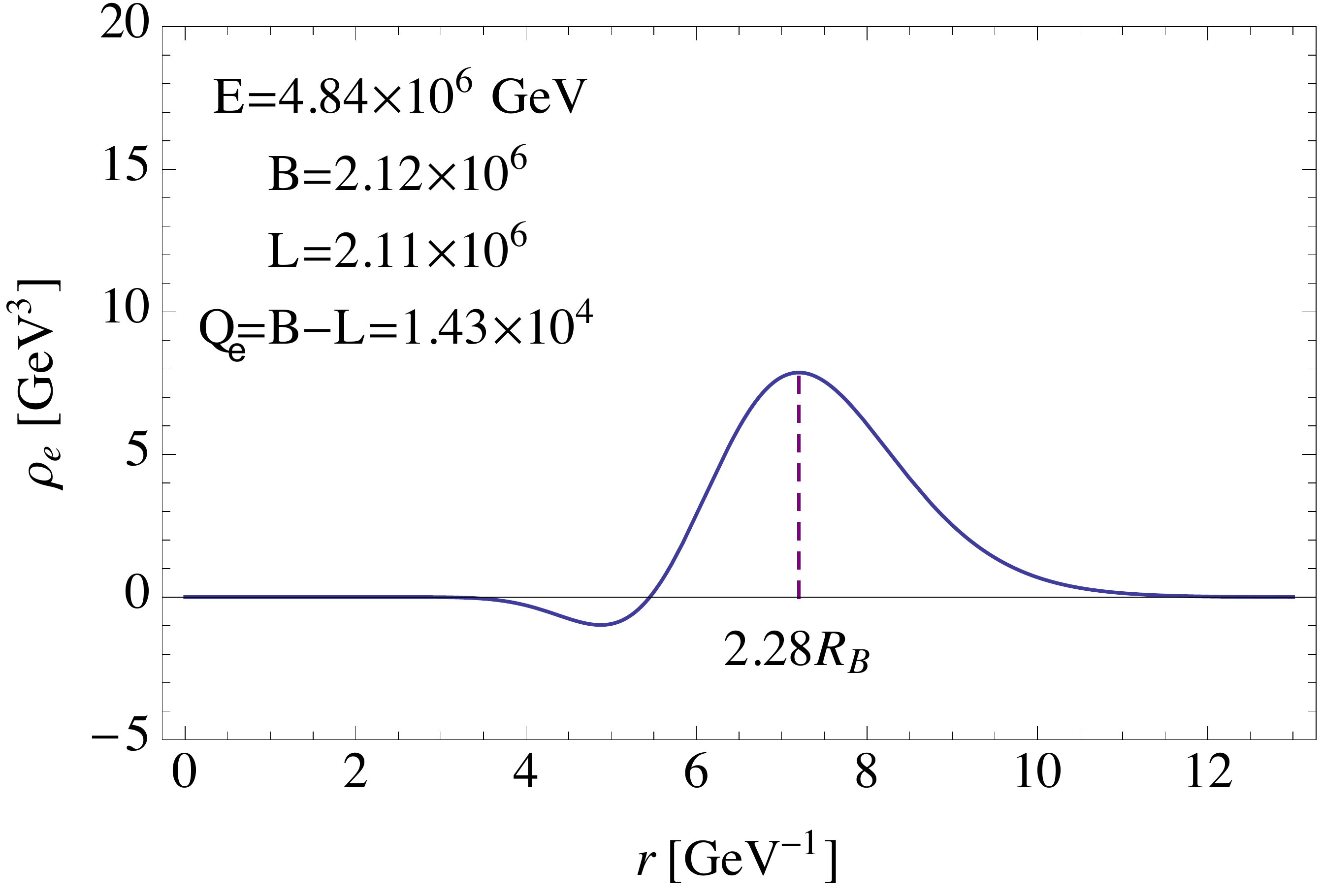}
 \end{minipage}
 \caption{Electric charge distributions of two scalar charged Q-balls with small nonzero electric charge for gauge mediation type charged Q-ball, with $m_\phi=1\text{GeV}$, and $e^2=0.002$~(Left), and new-type charged Q-ball, with $m_{3/2}=1\text{GeV}$, $|K|=0.1$, and $e^2=0.002$~(Right). We can see that the charge mainly distributes in the outer region, which, for new-type charged Q-ball, is more outer than the standard deviation $R_B\simeq R_L$.}
\label{fig:re}
\end{figure}

As reviewed in the previous section, there are simple expressions for energy, size, etc. of neutral Q-ball, but for charged Q-ball, the properties become difficult to analyze. However, in Ref.~\cite{hyk}, we numerically found, for gauge mediation type Q-ball, that the energy per unit charge is written as the neutral Q-ball expression plus coulomb potential at the surface of the Q-ball if the profile is not deformed very much\footnote{In Ref.~\cite{hyk}, we found that the realistic $Q_e$ is very small compared to $B$ or $L$, the deformation due to which is also small.}, as shown in Fig.~\ref{fig:prof}: 
\begin{align}
\left(\frac{\del E}{\del B}\right)_L&\simeq\left(\frac{\del E}{\del B}\right)_{L,\text{neutral}}+\frac{e^2Q_e}{4\pi R},\\
\left(\frac{\del E}{\del L}\right)_B&\simeq\left(\frac{\del E}{\del L}\right)_{B,\text{neutral}}-\frac{e^2Q_e}{4\pi R},
\label{eq:st}
\end{align}
where $R$ denotes $R_B\simeq R_L$.
The energy per unit charge is equal to the energy of a particle emitted from the Q-ball through the decay, we can interpret Eq.~(\ref{eq:st}) that the energy of a particle emitted from the charged Q-ball is written as the energy coming from the residence inside of the Q-ball, plus the coulomb potential at the surface. This interpretation is consistent with the fact that the electric charge density mainly distributes at the surface of the Q-ball as shown in Fig.~\ref{fig:re}, which means that the charged particles are emitted from the surface.  

\begin{figure}[t]
\centering
  \includegraphics[width=0.75\linewidth]{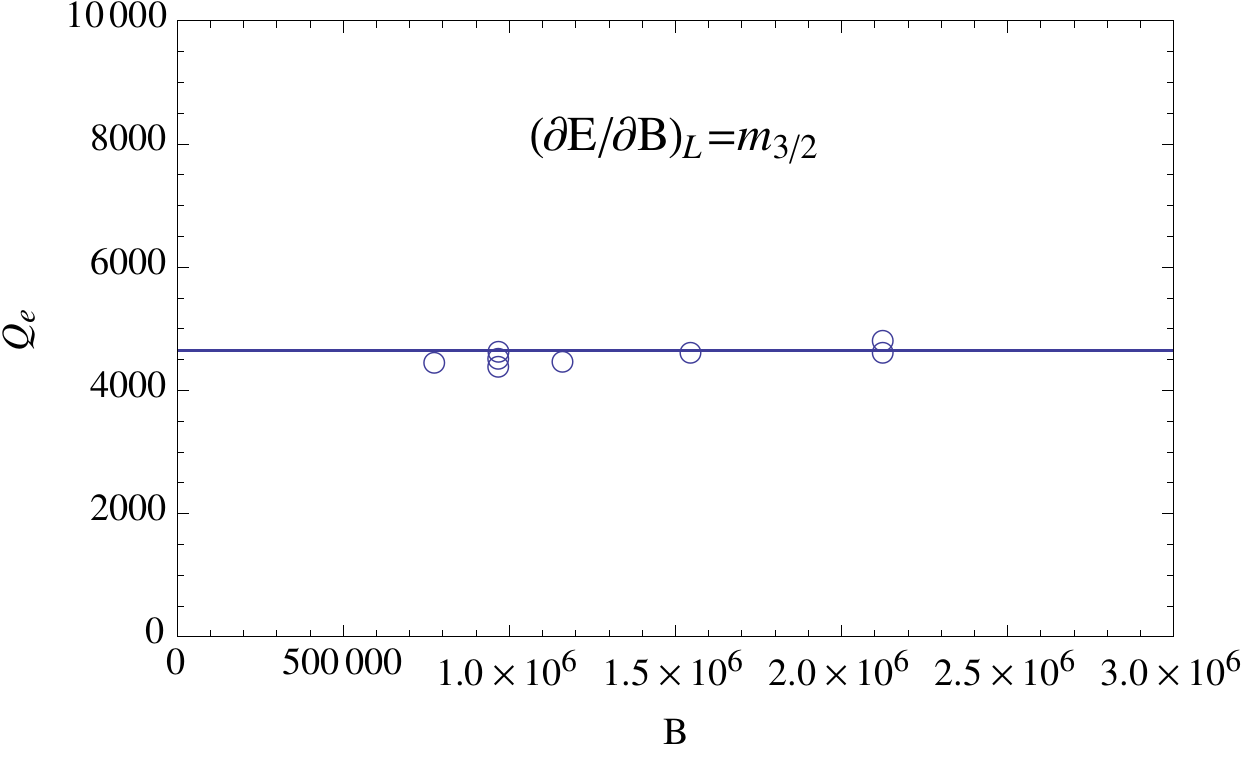}
 \caption{$B$-$Q_e$ plot for $\left(\del E/\del B\right)_L=m_{3/2}$. The dots are numerical result while the solid line indicates the analytic formula Eq.~(\ref{eq:st}) with $R$ replaced by $\tilde{R}$. We set $m_{3/2}=1\text{GeV}$, $|K|=0.1$, and $e^2=0.002$.}
\label{fig:st}
\end{figure}

In the case of new-type Q-ball, however, the profile is gaussian so the surface of the Q-ball must be defined carefully\footnote{As we see from Eq.~(\ref{eq:gau}), $R$ is just defined to be the standard deviation.}, and we numerically found that the above expression does not hold when we naively regarded $R$ given by Eq.~(\ref{eq:sizq}) as the surface of the Q-ball. However, from the physical interpretation above, we can specify where the particles will be emitted from, and we can see from Fig.~\ref{fig:re}, that the electric charge is located at more outer region than $R$. We found that Eq.~(\ref{eq:st}) still holds roughly for new-type Q-ball as well, when $R$ is replaced by a larger value, which is shown in Fig.~\ref{fig:st}. The best fit is obtained when $R$ is replaced by $\tilde{R}\equiv2.5R$, and from now on we use Eq.~(\ref{eq:st}) with $R$ replaced by $\tilde{R}$. The plot is for $\left(\del E/\del B\right)_L=m_{3/2}$, $|K|=0.1$, and $e^2=0.002$, but we confirmed the formula with $\tilde{R}$ for several values in the range $0.05<|K|<0.1$, $0.4m_{3/2}<\left(\del E/\del B\right)_L<m_{3/2}$. From now on, we assume this analytic formula for new-type charged Q-ball is valid for all parameter range we are interested in, and use it from the next section.
\subsection{$Q_e^{(\text{max})}$}
\label{sec:qmx}
Here we discuss maximum electric charge to which the Q-ball can be charged up by the decay. First of all, the condition $\left(\del E/\del L\right)_{B,\text{neutral}}\simeq m_{3/2}>m_e$ must be satisfied for the decay to occur initially
.
As electric attraction grows, we expect that the emitted particle becomes bounded to the Q-ball, which means the energy of the emitted particle becomes smaller than the rest mass:
\begin{align}
m_e>\left(\frac{\del E}{\del L}\right)_B&=m_{3/2}-\frac{e^2Q_e}{4\pi \tilde{R}}\\
&=m_{3/2}-\frac{e^2Q_e}{4\pi}\frac1{2.5|K|^{-1/2}m_{3/2}^{-1}},
\end{align}
thus
\begin{align}
Q_e&>2.5\alpha^{-1}|K|^{-1/2}\left(1-\frac{m_e}{m_{3/2}}\right)\\
&\simeq2.5\alpha^{-1}|K|^{-1/2},
\end{align}
where we used Eq.~(\ref{eq:st}) with $R$ replaced by $\tilde{R}$, as mentioned in the previous section.
As $Q_e$ grows further, the Bohr radius becomes smaller, eventually than the Q-ball size. If we naively assume that the particle is absorbed into the Q-ball again in that case
, the electric charge ceases to grow. The upper bound on the electric charge can be obtained as follows.  
\begin{align}
\frac{4\pi}{e^2Q_em_e}>\tilde{R}=2.5|K|^{-1/2}m_{3/2}^{-1},
\end{align}
thus
\begin{align}
Q_e<2.5^{-1}\alpha^{-1}|K|^{1/2}\frac{m_{3/2}}{m_e}.
\end{align}

Another effect which suppresses the growth of the electric charge is Schwinger effect, which states that if the electric field becomes large enough~($>E_{\text{schwinger}}\sim m_e^2/e$), pair creation occurs at interval of Compton length. Therefore, if the electric field at the surface of the Q-ball becomes larger than $E_{\text{schwinger}}$, the electron produced from pair creation is absorbed into the Q-ball
, suppressing the growth of the electric charge. The upper bound on the electric charge in this case is obtained as follows.
\begin{align}
E(\tilde{R})=\frac{eQ_e}{4\pi \tilde{R}^2}<E_{\text{schwinger}}=\frac{m_e^2}{e},
\end{align}
thus
\begin{align}
Q_e<(2.5)^2\alpha^{-1}|K|^{-1}\frac{m_e^2}{m_{3/2}^2}.
\end{align}

The above argument applies when the Q-ball size is larger than the electron Compton length. If the size of the Q-ball is smaller than the Compton length, the electric charge can grow further until the electric field at Compton length becomes $E_{\text{schwinger}}$, since the pair creation occurs at interval of Compton length, so we obtain 
\begin{align}
E(1/m_e)=\frac{eQ_e}{4\pi}m_e^2<E_{\text{schwinger}}=\frac{m_e^2}{e},
\end{align}
thus
\begin{align}
Q_e<\alpha^{-1},
\end{align}
that is, the fine structure constant inverse~($\simeq137$)
.

\begin{figure}[t]
\centering
  \includegraphics[width=0.75\linewidth]{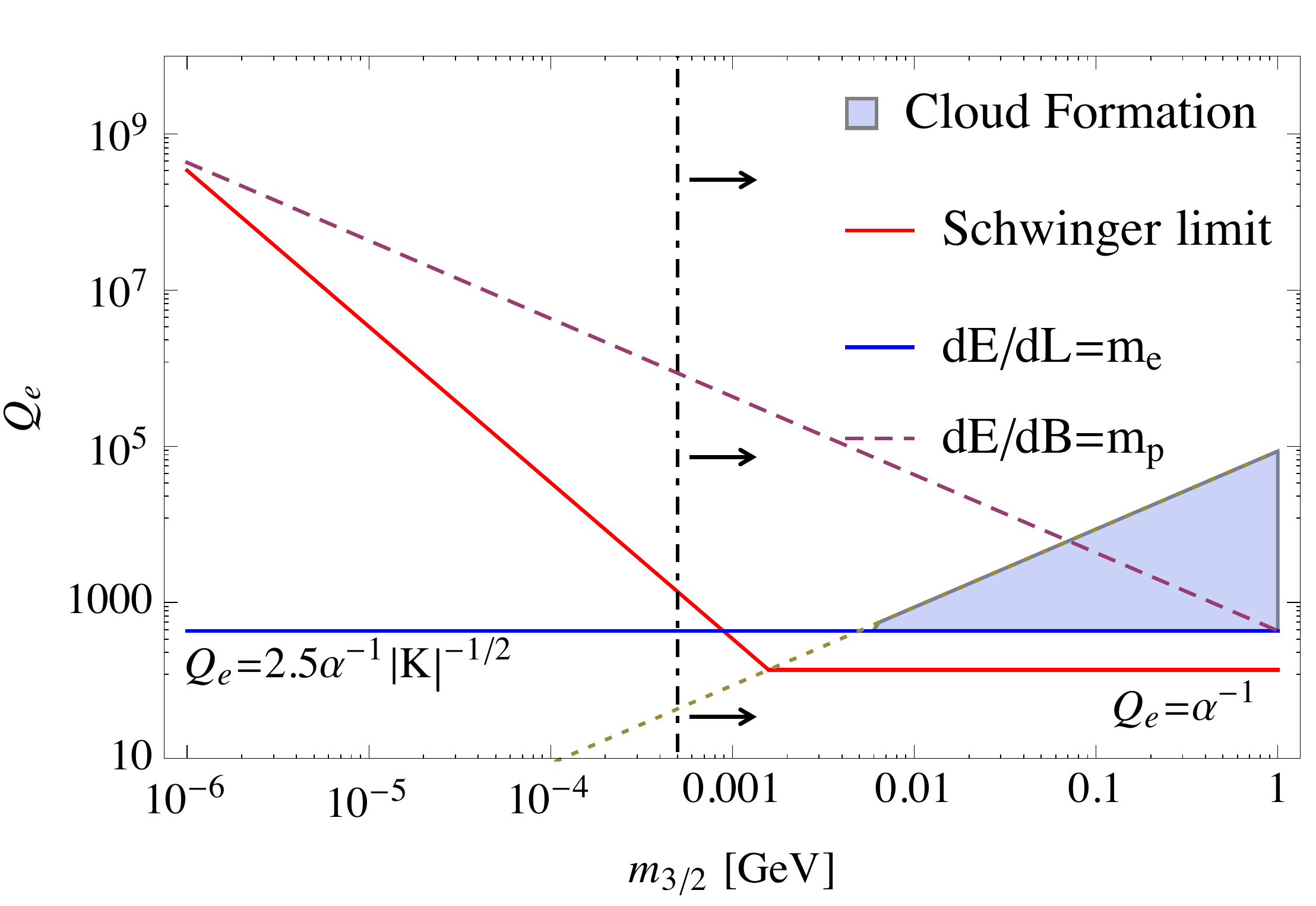}
 \caption{Upper bounds on electric charge of new-type charged Q-ball. The dotted line indicates when Q-ball size becomes equal to the Bohr radius, and the dashed-dotted line is $m_{3/2}=m_e$, whose right side is allowed. We have used $|K|=0.1$.}
\label{fig:qm}
\end{figure}

Finally, the electric repulsion may make the baryonic component unstable, so we must investigate whether the above upper bounds on the electric charge are still consistent with the stability condition. The baryonic component becomes unstable when the decay into nucleons becomes kinematically possible, which leads to the following stability condition:  
\begin{align}
m_p>\left(\frac{\del E}{\del B}\right)_L=m_{3/2}+\frac{e^2Q_e}{4\pi \tilde{R}},
\end{align}
thus
\begin{align}
Q_e&<2.5\alpha^{-1}|K|^{-1/2}\left(\frac{m_p}{m_{3/2}}-1\right)\\
&\simeq2.5\alpha^{-1}|K|^{-1/2}\frac{m_p}{m_{3/2}}.
\label{eq:bs}
\end{align}
Therefore, the growth of electric charge must stop before it reaches the upper bound given by Eq.~(\ref{eq:bs}).
In Fig.~\ref{fig:qm}, we illustrate the full upper bound on the electric charge of the Q-ball, and show that it safely maintain the stability of the baryonic component.
\section{Present relics and MICA constraint}
\label{sec:ev}
Here we discuss the present relics from the new-type charged Q-balls and their detections. We focus on the case that the Bohr radius is larger than the Q-ball size when maximally charged up, so that we can treat the charged Q-ball as ordinary nucleus with extremely heavy mass, which makes the analysis easier. We see from Fig.~\ref{fig:qm}, that this condition is given by
\begin{align} 
m_{3/2}>|K|^{-1/2}m_e,
\label{eq:boq}
\end{align}
and electric charge becomes $Q_e=\alpha^{-1}$. Then, the charged Q-balls captures the other charged particles like ordinary nucleus
, which was already analyzed in Ref.~\cite{hyk2}. There, we used Saha's equation to roughly estimate when the recombination starts, that is, when $n_{1S}/n_{\text{Q-ball}}\sim1$~($n_{1S}$ : number density of bound state of a Q-ball and an electron), and we found that the recombination temperature becomes $T_{\text{rec}}\simeq8.6\text{keV}$, if we use, as binding energy, the usual value of nucleus and electron. Similarly, we could find when the recombination finishes, that is, when the charged Q-ball completely neutralizes. Since binding energy decreases due to the screening effect of the orbiting electrons for large enough elements, the neutralization temperature necessarily becomes smaller than the usual proton-electron recombination temperature. Therefore, the Q-balls do not neutralize, since the free electrons are already captured by protons. Thus, we can conclude that, if charged Q-balls with $Q_e\sim\alpha^{-1}$ are formed well before the BBN epoch, the present relics become 
 $+O(1)$ ion-like extremely heavy objects. 
These relics must eventually account for dark matter of the universe,~i.e:
\begin{align}
\frac{\rho_{\text{Q-ball}}}s=\frac{\rho_{\text{DM}}}s\simeq4.4\times10^{-10}~\mathrm{GeV}
\label{eq:dm}
\end{align}
where $\rho_{\text{DM}}$ and $s$ are the dark matter energy density and entropy density in the present universe, respectively.

The 
ion-like relics experience electromagnetic processes. Thus, experiments sensitive to those processes are applicable, and non-detection of the processes gives upper bounds on the amount or flux of the objects. Various upper bounds on the flux of the charged Q-ball relics are obtained in Ref.~\cite{kensyutu}, and the most stringent comes from MICA experiment~\cite{mica}, 
\begin{align}
F\lesssim2.3\times10^{-20}~\mathrm{cm}^{-2}\text{s}^{-1}\text{sr}^{-1},
\end{align}
where they did not observed any trails of heavy 
ion-like object in $10^9$ years old ancient muscovite mica crystals.The constraint is severe, mainly due to the long detection time, which is essentially the age of the mica. The other features are discussed in Ref.~\cite{hyk2}, along with the application on the gauge mediation type Q-ball case. 
Since the Q-ball is so heavy that the orbiting particles virtually has no effect on total mass, dark matter flux is given by
\begin{align}
F&\simeq\frac{\rho_{\text{DM}\odot}}{M_Q}v
\end{align}
where $\rho_{\text{DM}\odot}$ denotes dark matter energy density near the solar system, and $v$ is the virial velocity of the Q-balls.
Therefore, using $\rho_{\text{DM}\odot}\sim0.3~\mathrm{GeV/cm^3}$ and $v\sim10^{-3}$, we obtain the following constraint on the mass of Q-ball:
\begin{align}
M_Q\gtrsim3.9\times10^{26}~\mathrm{GeV}.
\label{eq:micac}
\end{align} 
This is a severe constraint, which easily reaches the typical mass of the Q-ball, and it constrains $\phi_{\text{osc}}$ via Eq.~(\ref{eq:typicalcharge}). We can translate it into the condition on $m_{3/2}$ and the reheating temperature $T_{\text{RH}}$, using a relation of $\phi_{\text{osc}}$ and reheating temperature $T_{\text{RH}}$ given by
\begin{align}
\frac{\rho_{\text{DM}}}{s}\sim\frac{3T_{\text{RH}}}{4}\frac{M_Qn_\phi/Q}{3H_{\text{osc}}^2M_{\text{P}}^2}\sim\frac{9}{4}T_{\text{RH}}\frac{\phi_{\text{osc}}^2}{M_{\text{P}}^2},
\label{eq:rdm}
\end{align}
where $n_\phi=m_{\text{eff}}\phi_{\text{osc}}^2$, $3H_{\text{osc}}\simeq m_{\text{eff}}$, $m_{\text{eff}}\simeq m_{3/2}$, and $Q$ is global charge~($=B+L$ for the model considered in Sec.~\ref{sec:setp}) of the Q-ball. We used Eq.~(\ref{eq:mass})~\cite{21}. 
Inserting the observational value $\rho_{\text{DM}}/s\simeq4.4\times10^{-10}~\mathrm{GeV}$~\cite{rhos}, we obtain
\begin{align}
\phi_{\text{osc}}\simeq4.75\times10^{13}~\mathrm{GeV}\left(\frac{T_{\text{RH}}}{\mathrm{GeV}}\right)^{-1/2}.
\label{eq:phitrh}
\end{align}
Using this relation and Eq.~(\ref{eq:typicalcharge}), Eq.~(\ref{eq:micac}) is rewritten as
\begin{align}
T_{\text{RH}}\lesssim 5.79\times10^0\left(\frac{m_{3/2}}{\text{GeV}}\right)^{-1}~\mathrm{GeV},
\label{eq:micac2}
\end{align}
which is a condition on $m_{3/2}$ and $T_{\text{RH}}$. 

\begin{figure}[t]
\centering
  \includegraphics[width=0.85\linewidth]{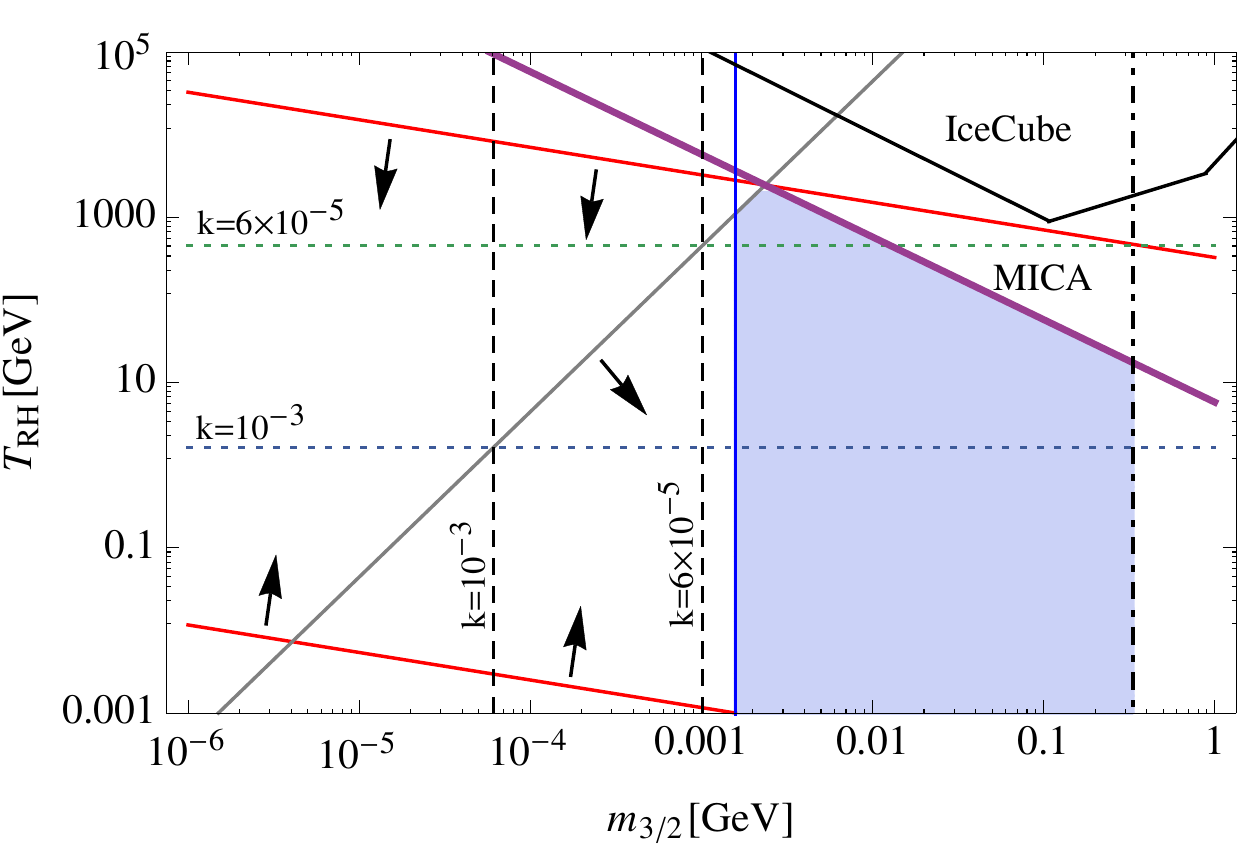}
 \caption{Allowed region for the new-type charged Q-ball as the dark matter~(shaded region). The thick line shows the upper bound Eq.~(\ref{eq:micac}), the dotted lines denote the lower bound Eq.~(\ref{eq:gad2}) for each value of $k$ shown, the dashed line corresponds to the $\Lambda_{\text{mess}}$-limit Eq.~(\ref{eq:h3}) with $g=1$, for each value of $k$ shown, the dashed-dotted line is the upper bound Eq.~(\ref{eq:bs}) for $Q_e=\alpha^{-1}$, and the blue line represents the lower bound Eq.~(\ref{eq:boq}). The bounds from Affleck-Dine mechanism are illustrated by red lines, and arrows are directed towards allowed region. The IceCube constraint is also represented at the upper right~\cite{icecube}.}
\label{fig:mftr}
\end{figure}

In Fig.~\ref{fig:mftr}, we show the allowed parameter region of $m_{3/2}$ and $T_{\text{RH}}$, using the analogous method to that in Ref.~\cite{icecube}, where the authors applied the IceCube constraint on the neutral Q-ball dark matter.  
The MICA constraint Eq.~(\ref{eq:micac2}) corresponds to the magenta solid line in the figure. 

The horizontal dotted lines indicate the condition that the gravity mediation effect dominates the potential of the Q-ball, which is given by Eq.~(\ref{eq:gad}).
Using Eq.~(\ref{eq:phitrh}), the condition is written as 
\begin{align}
T_{\text{RH}}\lesssim1.63\times10^{-6}~\mathrm{GeV}g^{-2}k^{-2}.
\label{eq:gad2}
\end{align}
In Fig.~\ref{fig:mftr}, the upper bounds on $T_{\text{RH}}$ for $k=6\times10^{-5}$ and $10^{-3}$ are shown. 

Another condition on gravitino mass~(SUSY breaking scale) comes from the observation value of Higgs boson mass at around 126~GeV~\cite{mgm,ft}:
\begin{align}
\Lambda_{\text{mess}}\equiv\frac{kF}{M_{\text{mess}}}\gtrsim5\times10^5~\mathrm{GeV}.
\label{eq:higgs}
\end{align}
Since the SUSY breaking scale is typically assumed to be small compared to the messenger mass
\begin{align}
kF<M_{\text{mess}}^2,
\end{align}
Eq.~(\ref{eq:higgs}) becomes
\begin{align}
\sqrt{kF}\gtrsim5\times10^5~\mathrm{GeV}.
\label{eq:h2}
\end{align}
Then, using Eq.~(\ref{eq:grma}), it reduces to 
\begin{align}
m_{3/2}\gtrsim6.1\times10^{-8}k^{-1}~\mathrm{GeV},
\label{eq:h3}
\end{align}
which corresponds to the vertical dashed line in Fig.~\ref{fig:mftr}.

The dashed-dotted line in Fig.~\ref{fig:mftr} indicates stability condition Eq.~(\ref{eq:bs}) for $Q_e=\alpha^{-1}$.
Finally, as mentioned in the beginning of the section, since we are considering the Q-ball smaller than the Bohr radius so that the potential the external particles experience can be approximated into Coulomb-type potential, Eq.~(\ref{eq:boq}) must be satisfied, which corresponds to the thin line in Fig.~\ref{fig:mftr}. 
The IceCube constraints, which are relevant only to the KKST process, are shown for comparison, and we see that more stringent MICA constraint makes the allowed region smaller.

By far, we treated the AD field at the onset of oscillation $\phi_{\text{osc}}$ as a free parameter. However, since the dynamics of AD field is essentially governed by the balance between the negative Hubble induced mass term and higher dimensional operator in superpotential $W\sim\phi^n/M^{n-3}$, $\phi_{\text{osc}}$ is determined as
\begin{align}
\phi_{\text{osc}}&\sim \left(H_{\text{osc}}M^{n-3}\right)^{1/(n-2)}\\
&\sim\left(m_{\text{eff}}M^{n-3}\right)^{1/(n-2)}\\
&=\left(m_{3/2}M^{n-3}\right)^{1/(n-2)},
\label{eq:str}
\end{align}
where we used $H_{\text{osc}}\sim m_{\text{eff}}=m_{3/2}$, due to the gravity mediation domination in the AD potential. Thus, we investigate whether the AD mechanism actually can generate the viable amplitude of the AD field, so that it is consistent with the constraints discussed above. In specific, we examine whether the reheating temperature determined by Eq.~(\ref{eq:phitrh}) and Eq.~(\ref{eq:str})
\begin{align}
T_{\text{RH}}\sim2.26\times10^{27}~\text{GeV}\left(\frac{m_{3/2}}{\text{GeV}}\right)^{-2/(n-2)}\left(\frac{M}{\text{GeV}}\right)^{-2(n-3)/(n-2)}
\label{eq:gos}
\end{align} 
is consistent with the allowed region in Fig.~\ref{fig:mftr}, for viable $n$ and $M$.  Since, we are considering $u^cu^cd^ce^c$ instead of usual $u^cd^cd^c$, $n$ becomes different from when we consider the neutral Q-ball. We take $n=8$, since $n=4$ is unfavored from proton decay
, while for $u^cd^cd^c$, we must take $n=6$ for instance. Then, we find that Eq.~(\ref{eq:gos}) is consistent with the allowed region for $4.16\times10^{-4}M_{\text{P}}<M<3.07M_{\text{P}}$, which is a reasonable range for $M$, whose bounds are illustrated by red lines in Fig.~\ref{fig:mftr}.

\section{Conclusions and discussion}
\label{sec:conc}
In gauge mediation models, gravity mediation effect can still dominate over the gauge mediation effect in the scalar potential if the field value is large. In that case, new-type Q-balls, which have different properties from gauge mediation type Q-balls, are formed after the Affleck-Dine baryogenesis. New-type Q-ball has property that it is stable against the decay into baryons but it can decay into leptons lighter than gravitino. Therefore, new-type Q-balls that carry both baryon and lepton charges can be electrically charged due to the leptonic decay only, which are called charged or gauged Q-balls~\cite{gaugedqball}. Also, they are stable by virtue of the stability of the baryonic component, thus, the new-type charged Q-ball can be a viable candidate for dark matter in the present universe. In this paper, we examined the allowed parameter region for new-type charged Q-ball dark matter. We focused on the case where the charged Q-ball can be treated as ordinary nucleus, in order to simplify the analysis. Then, we found that electric charge of charged Q-balls becomes $Q_e\sim\alpha^{-1}$, and the present relics of them  become 
$+O(1)$ ion-like extremely heavy objects. The relics can be treated as ordinary 
ions, so they are detectable by MICA experiment, where no trail of heavy 
ion-like object is observed in $10^9$ years old ancient mica crystals. This gives the stringent constraint on the dark matter flux, since the detection time, which is the age of the mica crystals, is extremely long. We translated the constraint into that on the gravitino mass $m_{3/2}$ and reheating temperature $T_{\text{RH}}$, as done in Ref.~\cite{icecube} for the IceCube constraint on the neutral Q-ball dark matter. As a result, we found that the MICA constraint makes the allowed region in $m_{3/2}$ - $T_{\text{RH}}$ smaller.

Since the dynamics of AD field is essentially governed by the AD potential, $\phi_{\text{osc}}$ is actually not a free parameter, and determined by the balance between the negative Hubble induced mass term and higher dimensional operator in superpotential $W\sim\phi^n/M^{n-3}$. We examined whether the amplitude of the AD field allowed by the constraints in Fig.~\ref{fig:mftr} is actually generated. Since the realistic example of the flat direction in our case is $u^cu^cd^ce^c$, instead of usual $u^cd^cd^c$ for the AD baryogenesis, $n$ becomes different from the latter. We took $n=8$, since $n=4$ is unfavored from proton decay, while for $u^cd^cd^c$, $n=6$ is taken, for instance. We found the parameter range $4.16\times10^{-4}M_{\text{P}}<M<3.07M_{\text{P}}$ is consistent with the allowed region in Fig.~\ref{fig:mftr}, which is a viable range for $M$.  


We found that the flat direction which forms charged Q-balls can explain dark matter in the universe, but the number of baryonic particles emitted from evaporation of the charged Q-ball due to thermal bath is too small 
  to explain the baryon asymmetry in the universe~\cite{fkk}. However, by using another flat direction, the baryon asymmetry can be explained as well.   

We focused on the region where the charged Q-ball can be treated as ordinary nucleus, which means the Bohr radius is larger than the Q-ball size. For the opposite case, where the Bohr radius is smaller than the Q-ball size, we naively assumed in Sec.~\ref{sec:qmx}, that the emitted particle is absorbed again into the Q-ball, which suppresses the further growth of electric charge. However, the physics when the Bohr radius is smaller than the Q-ball size may be a complex issue. For example, the particle may orbit inside of the Q-ball, experiencing the different potential from Coulomb type. Then the bound state with different properties including binding energy may be formed, which will change the present relics. Then, the detections may have to be applied differently as well. It will be a future task to investigate what will happen when orbit is smaller than the Q-ball size. 

We assumed $u^cu^cd^ce^c$-like $B$ and $L$ flat direction which includes only electron as the leptonic component. However, if we consider the direction which consists of neutrino component, for example $QQQL$, the scenario may be different by far. For instance, due to the decay channel into neutrinos, the Q-ball may have $SU(2)$ charge, and a fundamental theory of non-abelian gauged Q-ball may need to be considered. 

\vspace{1cm}

\section*{Acknowledgments}
J.H. would like to thank Hye-Sung Lee for helpful comments. This work is supported by Grant-in-Aid for Scientific Research from the Ministry of Education, Science, Sports, and Culture (MEXT), Japan, No. 15H05889 and No. 25400248 (M.K.). The work is also 
supported by World Premier International Research Center Initiative (WPI Initiative), MEXT, Japan. 

\vspace{1cm}



\end{document}